\documentclass{IET-Conf-Paper}

\usepackage{amsmath, amssymb, amsfonts, amsthm, booktabs, cancel, comment, eurosym, float, graphicx, mathtools, multirow, nicefrac, placeins, soul, steinmetz, stfloats, siunitx, svg, url, xcolor, tabularx}%
\usepackage{hyperref}
\usepackage{nomencl}
\usepackage{cleveref}
\makenomenclature

\usepackage{pifont}

\usepackage{etoolbox}
\renewcommand\nomgroup[1]{%
\item[\bfseries
  \ifstrequal{#1}{I}{\textit{Sets and Indices}}{%
  \ifstrequal{#1}{P}{\textit{Parameters}}{%
  \ifstrequal{#1}{A}{\textit{Acronyms}}{%
  \ifstrequal{#1}{W}{\textit{Dual variables}}{%
  \ifstrequal{#1}{V}{\textit{Variables}}{}}}}}%
]}

\floatstyle{ruled}
\newfloat{model}{thp}{lop}
\floatname{model}{\textsc{Model}}

\DeclareMathOperator*{\minimize}{min.}
\DeclareMathOperator*{\maximize}{max.}

\newcommand{\D}{\mathcal{D}}
\newcommand{\Q}{\mathcal{Q}}
\newcommand{\E}{\mathcal{E}}
\newcommand{\T}{\mathcal{T}}

\DeclareSIUnit\h{h}
\DeclareSIUnit\A{A}
\DeclareSIUnit\Ah{Ah}
\DeclareSIUnit\MWh{MWh}
\DeclareSIUnit\kWh{kWh}
\DeclareSIUnit\MW{MW}
\DeclareSIUnit\MVAr{MVAr}
\DeclareSIUnit\MVA{MVA}
\DeclareSIUnit\pu{p.u.}
\DeclareSIUnit\EurokWh{\text{\euro}/kWh}
\DeclareSIUnit\USDkWh{\text{\$}/kWh}
\DeclareSIUnit\USDMW{\text{\$}/MW}
\DeclareSIUnit\USDMWh{\text{\$}/MWh}
\DeclareSIUnit\USDkWhpar{\text{\$}/{kWh^*}}
\DeclareSIUnit\USDMWpar{\text{\$}/{MW^*}}
\DeclareSIUnit\EuroMWh{\text{\euro}/MWh}

\usepackage{mleftright}
\mleftright

\let\originalleft\left
\let\originalright\right
\renewcommand{\left}{\mathopen{}\mathclose\bgroup\originalleft}
\renewcommand{\right}{\aftergroup\egroup\originalright}

\newcolumntype{P}[1]{>{\centering\arraybackslash}p{#1}}

\usepackage{threeparttable}

\usepackage{subfigure}

\usepackage{algorithmic}
\usepackage{graphicx}
\usepackage{textcomp}
\usepackage{xcolor}
\usepackage{IEEEtrantools}

\begin{document}

\title{Pricing in Integrated Heat and Power Markets}

\author{Alvaro Gonzalez-Castellanos,\ad{1}\corr, David Pozo\ad{1}, Aldo Bischi \ad{1}}

\address{\add{1}{Center for Energy Science and Technology, Skolkovo Institute of Science and Tehcnology, Moscow, Russia}
\email{alvaro.gonzalez@skolkovotech.ru}}

\keywords{INTEGRATED HEAT AND POWER MARKET, COMBINED HEAT AND POWER, COST RECOVERY, CONVEX OPTIMIZATION}

\begin{abstract}
There is a growing interest in the integration of energy infrastructures to increase systems' flexibility and reduce operational costs.
The most studied case is the synergy between electric and heating networks.
Even though integrated heat and power markets can be described by a convex optimization problem, prices derived from dual values do not guarantee cost recovery.
In this work, a two-step approach is presented for the calculation of the optimal energy dispatch and prices.
The proposed methodology guarantees cost-recovery for each of the energy vectors and revenue-adequacy for the integrated market.
\end{abstract}

\maketitle

\nomenclature[V]{$m^\text{p}_g$}{Marginal electricity generation cost. \hfill [\SI{}\USDMWh{}]}
\nomenclature[V]{$m^\text{h}_g$}{Marginal heat generation cost. \hfill [\SI{}\USDMWh{}]}
\nomenclature[V]{$u^\text{pd}_i, u^\text{cd}_i$}{Electric demand uplift payments and charges. \textcolor{white}{Just filling some space}\hfill [\SI{}\USDMWh{}]}
\nomenclature[V]{$u^\text{pg}_g, u^\text{cg}_g$}{Electric generation uplift payments and charges. \hfill [\SI{}\USDMWh{}]}
\nomenclature[V]{$v^\text{pd}_j, v^\text{cd}_j$}{Heating demand uplift payments and charges. \textcolor{white}{Just filling some space}\hfill [\SI{}\USDMWh{}]}
\nomenclature[V]{$v^\text{pg}_g, v^\text{cg}_g$}{Heating generation uplift payments and charges. \hfill [\SI{}\USDMWh{}]\\}
\nomenclature[V]{$\Phi_i$}{Net electric demand value. \hfill [\$]}
\nomenclature[V]{$\Psi_i$}{Net heating demand value. \hfill [\$]}
\nomenclature[V]{$\Pi_g$}{Electric generation profit. \hfill [\$]}
\nomenclature[V]{$\Theta_g$}{Heating generation profit. \hfill [\$]}
\nomenclature[A]{PM--IHPM}{Pricing method for the integrated heat and power market.\\}
\nomenclature[A]{IHPD}{Integrated heat and power dispatch.}
\nomenclature[A]{IHPM}{Integrated heat and power market.}
\nomenclature[A]{MO}{Market operator.}
\nomenclature[W]{$\lambda$}{Electricity price of the IHPD. \hfill [\SI{}\USDMWh{}]}
\nomenclature[W]{$\gamma$}{Heating price of the IHPD. \hfill [\SI{}\USDMWh{}]}
\nomenclature[W]{$\mu_{g,l}$}{Dual variable related to the operation at the generation boundaries. \hfill [\$]}
\nomenclature[V]{$\lambda^\text{PM}$}{Electricity price after correction. \hfill [\SI{}\USDMWh{}]}
\nomenclature[V]{$\gamma^\text{PM}$}{Heating price after correction. \hfill [\SI{}\USDMWh{}]}
\nomenclature[V]{$d_i$}{Electric demand. \hfill [MWh]}
\nomenclature[V]{$q_j$}{Heating demand. \hfill [MWh]}
\nomenclature[V]{$p_g, h_g$}{Electricity and heat generation. \hfill [MWh]}
\nomenclature[P]{$K_{g,l}^{(\cdot)}$}{Coefficients for generation boundaries. \hfill \\
}
\nomenclature[P]{$c_{(\cdot),g}^{(\cdot)}$}{Generation cost coefficients. \hfill }
\nomenclature[P]{$b_i^\text{p}$}{Value of electric consumption. \hfill [\SI{}\USDMWh{}]}
\nomenclature[P]{$b_i^\text{h}$}{Value of heating consumption. \hfill [\SI{}\USDMWh{}]}
\nomenclature[I]{$i \in \D$}{Electricity user.}
\nomenclature[I]{$j \in \Q$}{Heating user.}
\nomenclature[I]{$g \in \E$}{Electric energy generation units.}
\nomenclature[I]{$g \in \T$}{Thermal energy generation units.}
\nomenclature[I]{$i \in D^+$}{Dispatched electricity user.}
\nomenclature[I]{$j \in \Q^+$}{Dispatched heating user.}
\nomenclature[I]{$g \in \E^+$}{Dispatched electric energy generation units.}
\nomenclature[I]{$g \in \T^+$}{Dispatched thermal energy generation units.}
\nomenclature[I]{$l$}{Operation bounds for generation units.\\}
\printnomenclature[0.8in]

\section{Introduction}
The increasing penetration of renewable energy and the decentralization of energy systems operations requires the enhancement of the systems' flexibility.
Operational integration of energy systems allows increasing flexibility through the use of existing synergies.
The use of cogeneration units, electric boilers, and heat pumps in district heating systems links the heat and electricity systems \cite{Averfalk2017}, providing an opportunity to improve the systems' flexibility and economic performance through their integration \cite{Guelpa2019}.

The operation integration of heat and power systems at the urban level can be done with an integrated heat and power market (IHPM). 
An IHPM is an energy market in which electricity and heat generation is co-optimized.
In an IHPM, the market operator (MO) collects the generators' cost bids and operating regions, as well as the users' price and quantity bids.
The MO then performs a dispatch (IHPD) to optimally schedule the generation and demand, maximizing the system's social welfare, i.e., the sum of the consumers' surplus and the generators' profits.
The operation integration of heat and power systems has been widely studied in the literature; a comprehensive review can be found in \cite{Guelpa2019}. 
From a market perspective, the integration of heat and electricity systems has been analyzed through dispatch coordination \cite{Mitridati2020, Cao2019}, energy trading \cite{Chen2019}, and in entirely integrated markets \cite{Schwele2020, Deng2019}.

In deregulated electricity and heat markets, it is common practice to set energy prices based on duality theory, i.e., the price represents the cost of generating an additional unit of energy. 
However, in IHPMs, the use of marginal energy pricing does not guarantee cost recovery for cogeneration units, i.e., the energy payment does not cover the related generation cost. 
In \cite{Deng2019}, the authors identified the possibility of failure to provide cost recovery when cogeneration units operate along the boundaries of the feasible operation region.
Failure to recover generation costs is exemplified in heat- and electricity-oriented operation modes for cogeneration plants. 
However, a methodology to modify the energy prices to guarantee cost recovery was not proposed. 
To the authors' best knowledge, it was not addressed in other works either.

Given the growing interest in designing new integrated heat and power markets, this letter presents a linear programming methodology for energy pricing in them. 
The energy prices are calculated after the optimal dispatch is obtained and guarantee the cost recovery on each of the energy vectors, as well as the market's revenue adequacy.
For the sake of simplicity and to properly present the properties of the proposed method, the pricing procedure is performed in an integrated heat and power system without temporal or network constraints; which can be easily included in the proposed methodology.
The benefits of deriving the energy prices with the presented approach are exemplified with a numerical test case in which two typical load scenarios, summer and winter, are portrayed. 
Additionally, the need to guarantee cost recovery for both the produced electricity and heat is demonstrated through a comparison with a formulation in which prices ensure net cost recovery for the cogenerators.
%
%
%

\section{Integrated Heat and Power Dispatch}
The objective of the IHPD is the optimal generation dispatch that maximizes the system's social welfare \eqref{Eq: Obj IHPD}.
In \eqref{Eq: Obj IHPD}, the dispatched electric demand is given by \(d_i\) and its offered price by \(b^\text{p}_i\). 
The dispatched heating demand is given by \(q_j\) and its price bid by \(b^\text{h}_j\).
The operational cost \(C_g\) of the dispatched generation is calculated as a second-degree polynomial of the generated electricity and heat, respectively, for generators \(g {\in} \E\) and \(g {\in} \T\).
The system's power and heat balances are given by \eqref{Eq: P Balance} and \eqref{Eq: H Balance}, respectively.
A common approach for the characterization of a cogeneration unit's operation region is through the convex combination of sampling points that result in a convex polyhedron \cite{Guo1996}. 
This operation region can be described in terms of its bounds \eqref{Eq: PHSampling} \cite{Deng2019}.
Expression \eqref{Eq: PHSampling} allows to characterize electric-only generation units by setting \(K^\text{h}_l{=}0, ~\forall l\). Heat-only units with \(K^\text{p}_l{=}0, ~\forall l\).
The electric and heat demand limits are set by \eqref{Eq: dFlex} and \eqref{Eq: qFlex}.
%
\begin{model}[ht]
\caption{Integrated Heat \& Power Dispatch \hfill}
\label{IHPD}
\begin{subequations} 
\label{Mod: IHPD}
\vspace{-2\jot}
\begin{IEEEeqnarray}{lll}
    \IEEEeqnarraymulticol{3}{c}{ \maximize \qquad SW = \smashoperator{\sum_{i \in \mathcal{D}}} b^\text{p}_id_i + \smashoperator{\sum_{j \in \mathcal{Q}}} b^\text{h}_jq_j - \smashoperator{\sum_{g \in \mathcal{E} \cup \mathcal{T}}} C_g} \label{Eq: Obj IHPD} \IEEEyesnumber\\
    \IEEEeqnarraymulticol{2}{l}{\text{subject to}} & \IEEEnonumber\\
    \IEEEeqnarraymulticol{2}{l}{C_g {=} c_{2,g}^\text{p}p_g^2 {+} c_{1,g}^\text{p}p_g {+} c_{2,g}^\text{h}h_g^2 {+} c_{1,g}^\text{h}h_g{+}c^\text{h,p}h_gp_g {+} c^0_{g}},  &\forall g \IEEEeqnarraynumspace \label{Eq: Gencosts} \\
    (\lambda^\text{}){:} ~ & \sum_{i \in \D} d_{i} - \sum_{g \in \E} p_{g} = 0, \label{Eq: P Balance} \\
    (\gamma^\text{}){:} ~ & \sum_{j \in \Q} q_{j} - \sum_{g \in \T} h_{g} = 0, \label{Eq: H Balance} \\
    (\mu^\text{}_{g,l}){:} \quad& p_g K^\text{p}_{g,l} + h_g K^\text{h}_{g,l} \leq K^\text{0}_{g,l}, \qquad \qquad \qquad &\forall g,l  \IEEEeqnarraynumspace \label{Eq: PHSampling}\\
    & 0 \leq d_i \leq \overline{d}_i, &\forall i \IEEEeqnarraynumspace\label{Eq: dFlex}\\
    & 0 \leq q_j \leq \overline{q}_j, &\forall j. \IEEEeqnarraynumspace\label{Eq: qFlex}
    \vspace{-2\jot}
\end{IEEEeqnarray}
\end{subequations}
\end{model}

The energy prices of the IHPD can be derived from the KKT stationarity conditions regarding \(p_g\) and \(h_g\):
\begin{IEEEeqnarray}{rCll}
    \lambda &=& m^\text{p}_g + \sum_l \mu_{g,l}K^\text{p}_{g,l},
    \quad & \forall g \in \E \IEEEyesnumber\IEEEyessubnumber*\IEEEeqnarraynumspace\label{Eq: Marginal p_a}\\
    \gamma &=& m^\text{h}_g + \sum_l \mu_{g,l}K^\text{h}_{g,l}, 
    \qquad & \forall g \in \T \IEEEeqnarraynumspace \label{Eq: Marginal h_a}
    \end{IEEEeqnarray}
where the marginal electricity and heat generation costs are given by \(m^\text{p}_g{=}{\partial{C_g}}/{\partial{p_g}}\) and \(m^\text{h}_g{=}{\partial{C_g}}/{\partial{h_g}}\), respectively.

\begin{figure}[!h]
    \centering
    \includegraphics[width=\columnwidth]{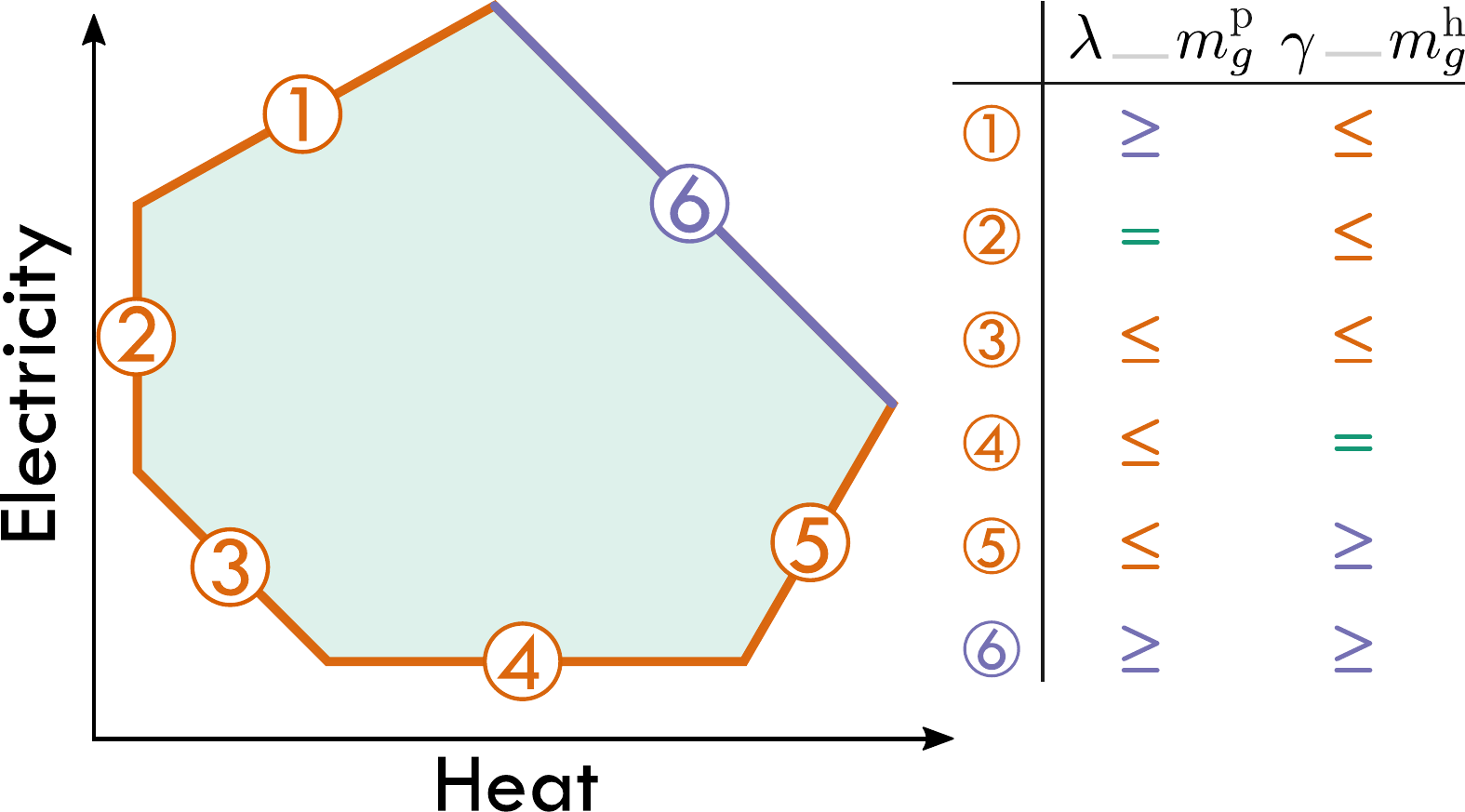}
    \caption{Relationship between energy prices and marginal generation costs along the bounds of a general operation region for cogenerators. 
    In the green area the energy prices equal the unit's marginal generation costs.
    Along the orange edges \ding{172}, \ding{173}, \ding{174}, \ding{175}, and \ding{176}, the energy price is lower than the associated marginal generation cost for electricity, heat or both. 
    Along the purple boundary \ding{177}, the energy prices are higher than the associated marginal generation costs.
    }
    \label{Fig: OpRegion}
\end{figure}

If a generation unit \(g\) is dispatched without having active any of its technical limits, i.e., in the green area of Fig. \ref{Fig: OpRegion}, then \(\mu_{g,l}{=}0, ~ \forall l\); \(\lambda {=} m^\text{p}_g\), and \(\gamma {=} m^\text{h}_g\). 
However, if the unit is dispatched at the generation bounds, i.e., the expression \eqref{Eq: PHSampling} becomes binding for any bound \(l\), the dual variable \(\mu_{g,l} {\ge} 0\) and the sign of \(\sum_l \mu_{g,l}K^\text{h/p}_{g,l}\) depends on the value of the active bound's coefficient. 
Therefore, depending on the generation bound on which the unit operates, the price of energy can become smaller than the associated marginal generation cost; for heat, electricity, or both.
More specifically: 
    i) \(\gamma {\le} m^\text{p}_g\) along bounds \ding{172}, \ding{173}, and \ding{174},
    and
    ii) \(\lambda {\le} m^\text{h}_g\) on \ding{174}, \ding{175}, and \ding{176}.
Note that the generator has positive profit along the boundary \ding{177}.
Consequently, despite the IHPD being a convex model, marginal prices do not guarantee cost recovery for cogenerators \cite{Deng2019}.
A similar analysis could be made for the heating profits of heat pumps. However, since the heat pumps consume electricity to produce heat, their operation region would fall on the fourth quadrant of Fig. \ref{Fig: OpRegion}.

\section{Pricing for the IHPD \label{Sec: Pricing}}
To guarantee cost recovery for both electricity and heat generation, we introduce a linear programming model for pricing in integrated heat and power markets, motivated by the dual pricing algorithm \cite{ONeill2017}. 
The pricing methodology (PM-IHPM) is presented in Model \ref{Pricing}, and its inputs are the optimal dispatched demand \(d_i^*\) and \(q_j^*\); optimal dispatched generation \(p_g^*\) and \(h_g^*\); and optimal electricity and heat prices \(\lambda^*\) and \(\gamma^*\) of the IHPD \eqref{Mod: IHPD}. 
In this manner, the pricing methodology presented in this section must be employed if after the IHPD it is detected that some generators do not recover their operation costs.
The sets of the dispatched electric and heat demands are denoted by \(\mathcal{D^+}\) and \(\mathcal{Q^+}\); and the dispatched electricity and heat generators by \(\mathcal{E^+}\) and \(\mathcal{T^+}\).

The objective of the PM-IHPM \eqref{Eq: Obj PM} is the minimization of the uplift payments made by the electric and heating demand (\(u^\text{pd}_i \text{~and~} v^\text{pd}_j\)), and generation (\(u^\text{pg}_g \text{~and~} v^\text{pg}_g\)).
The revenue neutrality of the electric and heat subsectors is respectively guaranteed by \eqref{Eq: Neutral P} and \eqref{Eq: Neutral H}.
The users' utility is defined in \eqref{Eq: Value d} and \eqref{Eq: Value q}, where \(u^\text{cd}_i \text{~and~} v^\text{cd}_j\) represent the demand charges.
\(\lambda^\text{PM}\) and \(\gamma^\text{PM}\) are the electricity and heat prices resultant of the PM-IHPM.
The generation utility is calculated in \eqref{Eq: Profit p} and \eqref{Eq: Profit h}.
The non-negativity of the users utility and generation profit is set by \eqref{Eq: non-confiscation i}--\eqref{Eq: non-confiscation g}, and the non-negativity of payments and charges by \eqref{Eq: nonNeg u-v}.

The pricing method \eqref{Mod: Pricing} could have multiple solutions. In addition, the monotonicity of the uplift charges is not guaranteed. 
Uplift charge monotonicity could be imposed in the PM-IHPMs by including logic constraints that state that greater charges should be made to agents with a higher surplus. For generators, the logical monotonicity constraint would be of the form
\(u_g^\text{pg} {\ge} u_k^\text{pg}\) if \(\Pi_g {\ge} \Pi_k,~ \forall g,k \in \E^+\); which can be translated into constraints with the use of binary variables.

\begin{model}[t]\caption{Pricing for the Integrated Heat \& Power Market\hfill}
\label{Pricing}
\begin{subequations} 
\label{Mod: Pricing}
\vspace{-2\jot}
\begin{IEEEeqnarray}{ll}
    \IEEEeqnarraymulticol{2}{c}{ \minimize \smashoperator{\sum_{i \in \mathcal{D}^+}} d_i^{*}u^\text{pd}_i 
    + \smashoperator{\sum_{j \in \mathcal{Q}^+}} q_j^{*} v^\text{pd}_j 
    + \smashoperator{\sum_{g \in \mathcal{E}^+}} p_g^{*}u_g^\text{pg} 
    + \smashoperator{\sum_{g \in \mathcal{T}^+}} h_g^{*}v_g^\text{pg}} \label{Eq: Obj PM} \IEEEeqnarraynumspace\IEEEyesnumber\\
    \IEEEeqnarraymulticol{2}{l}{\text{subject to}} \IEEEnonumber\\
    \IEEEeqnarraymulticol{2}{l}{\smashoperator{\sum_{i \in D^+}} d_{i}^{*}\left(u^\text{pd}_i-u^\text{cd}_i\right) + \smashoperator{\sum_{g \in \E^+}} p_{g}^{*}\left(u^\text{pg}_g-u^\text{cg}_g\right) = 0,} \label{Eq: Neutral P} \\
    \IEEEeqnarraymulticol{2}{l}{\smashoperator{\sum_{j \in \Q^+}} q_{j}^{*}\left(v^\text{pd}_j-v^\text{cd}_j\right) + \smashoperator{\sum_{g \in \T^+}} h_{g}^{*}\left(v^\text{pg}_g-v^\text{cg}_g\right) = 0, \quad} \label{Eq: Neutral H} \\
    \Psi_i = d_{i}^{*}\left(b_i^\text{p} - \lambda^\text{PM} + u^\text{pd}_i - u^\text{cd}_i\right) & \forall i \in \D^+ \label{Eq: Value d} \\
    \Phi_j = q_{j}^{*}\left(b_j^\text{h} - \gamma^\text{PM} + v^\text{pd}_j - v^\text{cd}_j\right) & \forall j \in \Q^+ \label{Eq: Value q} \\
    \Pi_g = p_{g}^{*}\left(\lambda^\text{PM} + u^\text{pg}_g - u^\text{cg}_g - m^\text{p}_g\right) & \forall g \in \E^+ \label{Eq: Profit p} \\
    \Theta_g = h_{g}^{*}\left(\gamma^\text{PM} + v^\text{pg}_g - v^\text{cg}_g - m^\text{h}_g\right) \hspace{1cm}& \forall g \in \T^+ \label{Eq: Profit h} \\
    0 \leq \Psi_i &\forall i \in D^+\label{Eq: non-confiscation i}\\
    0 \leq \Phi_j &\forall j \in Q^+\label{Eq: non-confiscation j}\\
    0 \leq \Pi_g, \Theta_g &\forall g \in \E^+ {\cup} \T^+ \label{Eq: non-confiscation g}\\
    0 \leq u^\text{pd}_i, u^\text{cd}_i, v^\text{pd}_j, v^\text{cd}_j, u^\text{pg}_g, u^\text{cg}_g, v^\text{pg}_g, v^\text{cg}_g. \label{Eq: nonNeg u-v}
    \vspace{-2\jot}
\end{IEEEeqnarray}
\end{subequations}
\end{model}

\section{Numerical Example}
A numerical example is devised to present the cost-recovery issues that can arise from an IHPD and the need to ensure separately cost-recovery for electricity and heat generation.
Two demand scenarios, summer and winter, are considered and presented in Table \ref{tab: Demand}. 
During the winter scenario, the maximum thermal demand and the offered bids increase considerably with respect to the summer values. 

\begin{figure*}[!h]
    \centering
    \subfigure[Electricity results]{
  		\includegraphics[width=0.9\columnwidth]{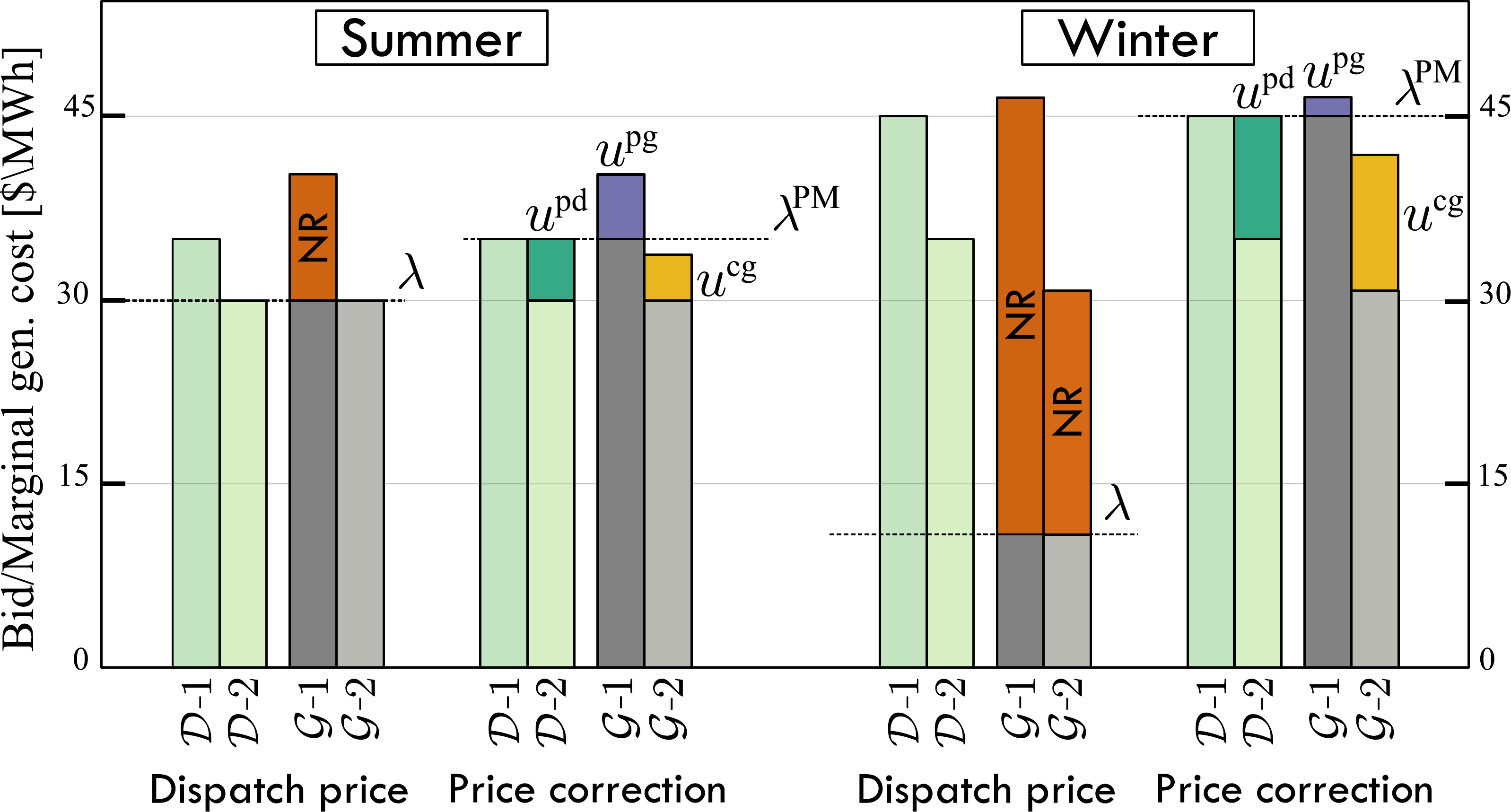}
  		\label{Fig: EResults}}\qquad\qquad
  	\subfigure[Heating results]{
  		\includegraphics[width=0.9\columnwidth]{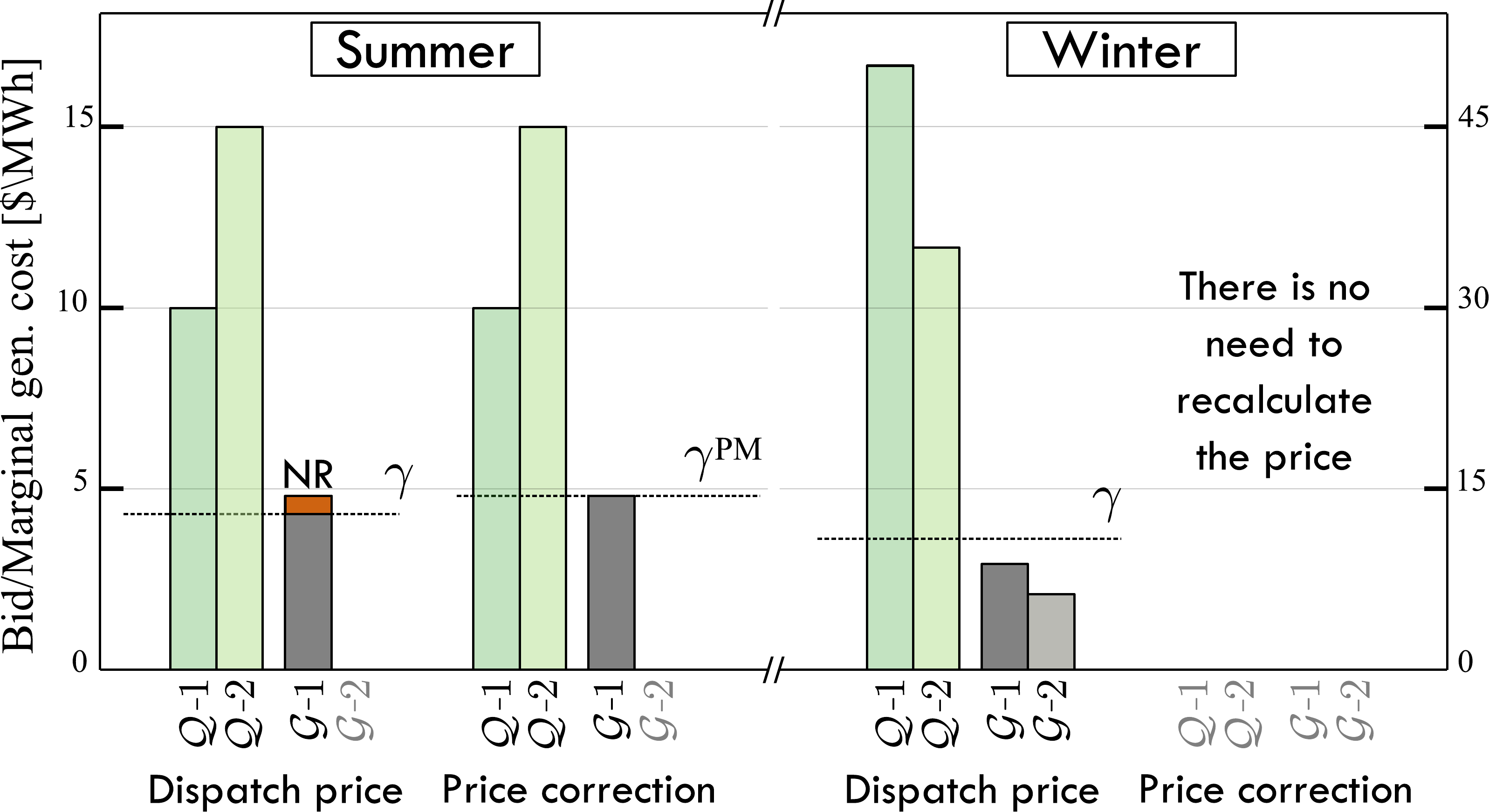}
  		\label{Fig: HResults}}
    \caption{Integrated heat and power dispatch, and pricing results for (a) electricity and (b) heating. NR = Non-recovered costs.\label{Fig: EHResults}}
\end{figure*}

\begin{figure}[t]
    \centering
    \includegraphics[width=0.8\columnwidth]{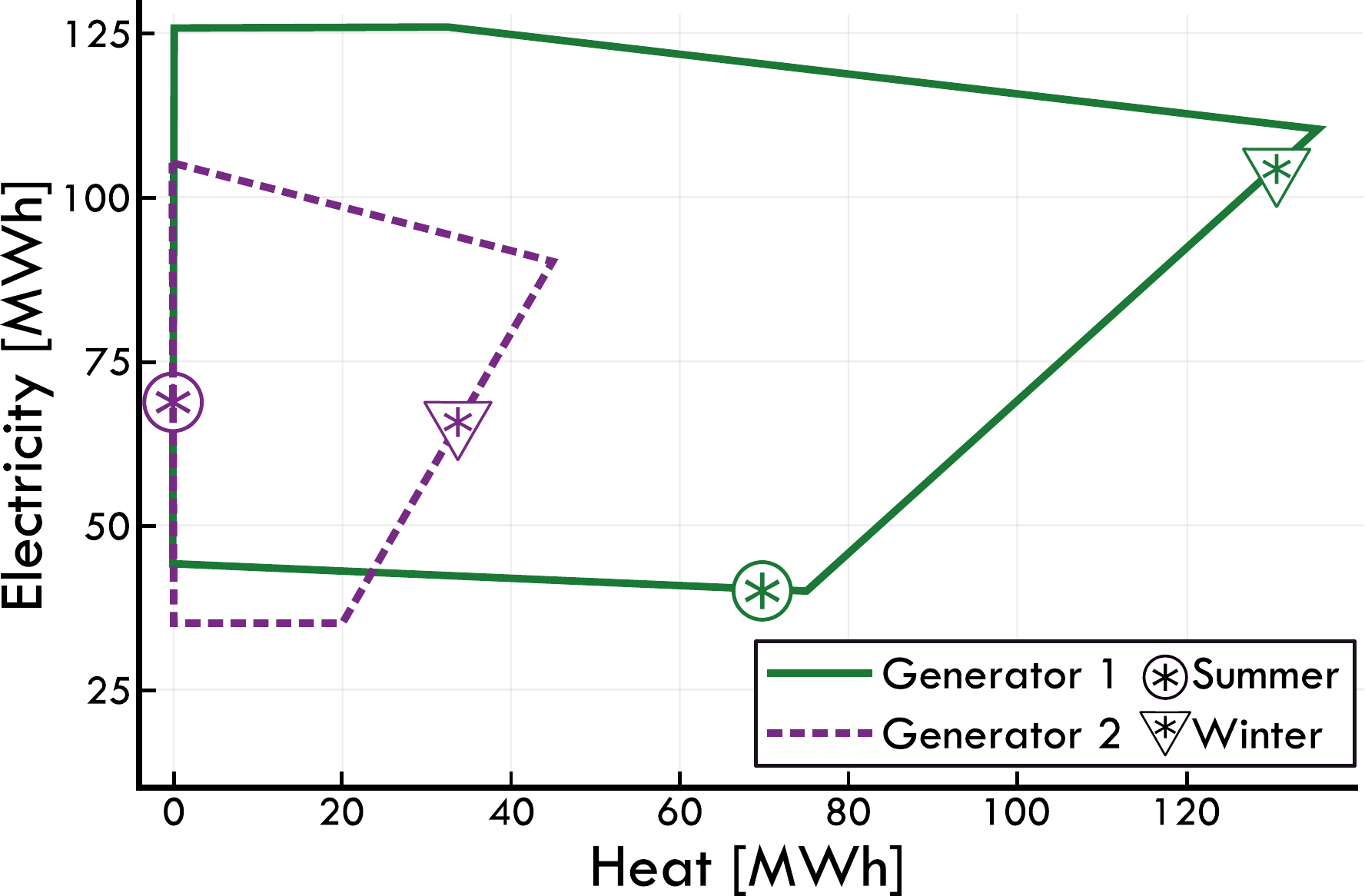}
    \caption{Scheduling results for the Summer and Winter cases.}
    \label{Fig: Results}
\end{figure}

The coefficients for the generation bounds are presented in Table \ref{Tab: GenData}, while Table \ref{Tab: GenCost} contains the cost coefficients. Fig. \ref{Fig: EHResults} presents the IHPD and the PM--IHPD results for the pricing of heat and electricity in both the summer and winter scenarios. 
The optimal scheduling points for the generation units in their operating regions are given in Fig. \ref{Fig: Results}.

%
%

\begin{table}[h]
\caption{Generation Boundaries' Coefficients\label{Tab: GenData}}
\centering
\begin{tabular}{@{}crrrrrr@{}}
\toprule
      & \multicolumn{3}{c}{Generator 1}                                                                                                & \multicolumn{3}{c}{Generator 2}                                                                                                \\
\(l\) & \multicolumn{1}{c}{\(K_{g,l}^\text{p}\)} & \multicolumn{1}{c}{\(K_{g,l}^\text{h}\)} & \multicolumn{1}{c}{\(K_{g,l}^\text{0}\)} & \multicolumn{1}{c}{\(K_{g,l}^\text{p}\)} & \multicolumn{1}{c}{\(K_{g,l}^\text{h}\)} & \multicolumn{1}{c}{\(K_{g,l}^\text{0}\)} \\
\midrule
1     & -1.00                                    & -0.05                                    & -44.00                                   & -1.00                                    & 0.00                                     & 35.00                                    \\
2     & -1.00                                    & 1.16                                     & 46.88                                    & -1.00                                    & 2.20                                     & 9.00                                     \\
3     & 1.00                                     & 0.15                                     & 130.70                                   & 1.00                                     & 0.33                                     & 105.00                                   \\
4     & 1.00                                     & 0.00                                     & 125.80                                   & 0.00                                     & -1.00                                    & 0.00                                     \\
5     & 0.00                                     & -1.00                                    & 0.00                                     & --                                       & --                                       & --                                       \\ 
\bottomrule
\end{tabular}
\end{table}
%

\begin{table}[h]
\caption{Generation Cost Coefficients \label{Tab: GenCost}}
\centering
\begin{tabular}{@{}ccccccc@{}}
\toprule
            & \(c_{2,g}^\text{p}\) & \(c_{1,g}^\text{p}\) & \(c_{2,g}^\text{h}\) & \(c_{1,g}^\text{h}\) & \(c_{g}^\text{hp}\) & \(c_{g}^0\) \\ \midrule
Generator 1 & 0.0435               & 36                   & 0.027                & 0.6                  & 0.011               & 12.5        \\
Generator 2 & 0.072                & 20                   & 0.02                 & 2.34                 & 0.04                & 15.65       \\ \bottomrule
\end{tabular}
\end{table}

%
\begin{table}[h]
\caption{Demand data \label{tab: Demand}}\centering
\resizebox{\linewidth}{!}{
\begin{tabular}{@{}cccccc@{}}
\toprule
\multirow{2}{*}{User} & \multirow{2}{*}{Type} & \multicolumn{2}{c}{Summer}                                                                                                                                          & \multicolumn{2}{c}{Winter}                                                                                                                                          \\ 
\cmidrule(l){3-6} 
                      &                       & \begin{tabular}[c]{@{}c@{}}Max. demand\\ \(\left[\text{MWh}\right]\)\end{tabular} & \begin{tabular}[c]{@{}c@{}}Bid\\ \(\left[\$/\text{MWh}\right]\)\end{tabular} & \begin{tabular}[c]{@{}c@{}}Max. demand\\ \(\left[\text{MWh}\right]\)\end{tabular} & \begin{tabular}[c]{@{}c@{}}Bid\\ \(\left[\$/\text{MWh}\right]\)\end{tabular} \\ \midrule
1                     & $\D$                     & 100                                                                                  & 35                                                                           & 100                                                                                  & 45                                                                           \\
2                     & $\D$                     & 70                                                                                   & 30                                                                           & 70                                                                                   & 35                                                                           \\
3                     & $\Q$                     & 60                                                                                   & 10                                                                           & 250                                                                                  & 50                                                                           \\
4                     & $\Q$                     & 10                                                                                   & 15                                                                           & 160                                                                                  & 45                                                                           \\
\bottomrule
\end{tabular}
}
\end{table}

\subsection{Summer Scenario \label{Sec: Summer}}
The results of the integrated heat and power dispatch for the summer scenario are given in Table \ref{Tab: IHPD Results_summer}. Table \ref{Tab: CA Results_summer} presents the results of the pricing methodology in the summer scenario.
For the summer scenario, the IHPD electricity and heat prices obtained from \eqref{Mod: IHPD} are, respectively, 30 and 4.27 \SI{}\USDMWh{}.
As seen in Fig. \ref{Fig: Results} and Fig. \ref{Fig: EHResults}, generator 1 does not recover its marginal electricity and heating costs (\(m^\text{p}_1 {=} 40.27~\SI{}\USDMWh{}\) and \(m^\text{h}_1 {=} 4.82~\SI{}\USDMWh{}\)); incurring in marginal generation deficits of 10.27 \SI{}\USDMWh{} and 0.55 \SI{}\USDMWh{}, respectively.

\begin{table}[t]
\caption{Summer IHPD Results\label{Tab: IHPD Results_summer}}\centering
\resizebox{\linewidth}{!}{
\begin{tabular}{@{}ccccc@{}}
\toprule
\multicolumn{5}{c}{\(SW\) = \$ 499.03 
~ \(\lambda^\text{}\) = 30 \SI{}\USDMWh{}            
~ \(\gamma^\text{}\) = 4.27 \SI{}\USDMWh{}} \\ \midrule
User      & Type                                                             & \begin{tabular}[c]{@{}c@{}}Dispatched\\ demand  \(\left[\SI{}\MWh{}\right]\)\end{tabular} & \begin{tabular}[c]{@{}c@{}}Electric\\ surplus  \(\left[\$\right]\)\end{tabular} & \begin{tabular}[c]{@{}c@{}}Heat\\ surplus \(\left[\$\right]\)\end{tabular} 
 \\ \midrule
1 & \(\D\) & 100 & 500 & -- \\
2 & \(\D\) & 9.71 & 0 & -- \\
3 & \(\Q\) & 60 & -- & 343.8 \\
4 & \(\Q\) & 10 & -- & 107.3 \\ \midrule 
Generator 
& \begin{tabular}[c]{@{}c@{}}Dispatched\\ electricity \(\left[\SI{}\MWh{}\right]\)\end{tabular} 
& \begin{tabular}[c]{@{}c@{}}Dispatched\\ heat \(\left[\SI{}\MWh{}\right]\)\end{tabular}   
& \begin{tabular}[c]{@{}c@{}}Electric\\ surplus  \(\left[\$\right]\)\end{tabular} 
& \begin{tabular}[c]{@{}c@{}}Heat\\ surplus  \(\left[\$\right]\)\end{tabular}
 \\ \midrule
1 & 40.27 & 70 & -413.57 & -38.5 \\
2 & 69.44 & 0 & 0 & 0  \\ \bottomrule
\end{tabular}
}
\end{table}
%
\begin{table}[t]
\caption{Summer PM--IHPD Results\label{Tab: CA Results_summer}}\centering
\resizebox{\linewidth}{!}{
\begin{tabular}{@{}ccccccc@{}}
\toprule
\multicolumn{7}{c}{\(SW\) = \$ 498.72 
~ \(\lambda^\text{PM}\) = 35 \SI{}\USDMWh{} 
~ \(\gamma^\text{PM}\) = 4.82 \SI{}\USDMWh{}} \\ \midrule
User      
& \begin{tabular}[c]{@{}c@{}}\(u^\text{pd}\)  \\ \(\left[\SI{}\USDMWh{}\right]\)\end{tabular}      
& \begin{tabular}[c]{@{}c@{}}\(u^\text{cd}\)  \\ \(\left[\SI{}\USDMWh{}\right]\)\end{tabular}      
& \begin{tabular}[c]{@{}c@{}}\(v^\text{pd}\)  \\ \(\left[\SI{}\USDMWh{}\right]\)\end{tabular}      
& \begin{tabular}[c]{@{}c@{}}\(v^\text{cd}\)  \\ \(\left[\SI{}\USDMWh{}\right]\)\end{tabular} 
& \begin{tabular}[c]{@{}c@{}}Electric\\ surplus \(\left[\$\right]\)\end{tabular}
& \begin{tabular}[c]{@{}c@{}}Heat\\ surplus \(\left[\$\right]\)\end{tabular} \\ \midrule
1 & 0 & 0 & -- & -- & 0 & -- \\
2 & 5 & 0 & -- & -- & 0 & -- \\
3 & -- & -- & 0 & 0 & --&310.62 \\
4 & -- & -- & 0 & 0 & --& 101.77\\ \midrule
Generator 
& \begin{tabular}[c]{@{}c@{}}\(u^\text{pg}\)  \\ \(\left[\SI{}\USDMWh{}\right]\)\end{tabular}      
& \begin{tabular}[c]{@{}c@{}}\(u^\text{cg}\)  \\ \(\left[\SI{}\USDMWh{}\right]\)\end{tabular}      
& \begin{tabular}[c]{@{}c@{}}\(v^\text{pg}\)  \\ \(\left[\SI{}\USDMWh{}\right]\)\end{tabular}      
& \begin{tabular}[c]{@{}c@{}}\(v^\text{cg}\)  \\ \(\left[\SI{}\USDMWh{}\right]\)\end{tabular} 
& \begin{tabular}[c]{@{}c@{}}Electric\\ surplus \(\left[\$\right]\)\end{tabular} 
& \begin{tabular}[c]{@{}c@{}}Heat\\ surplus \(\left[\$\right]\)\end{tabular} \\
1 & 5.27 & 0 & 0 & 0 & 0 & 0 \\
2 & 0 & 3.76 & 0 & 0 & 86.33 & 0 \\ \bottomrule
\end{tabular}
}
\end{table}

After the application of the PM-IHPM, the new electricity and heat prices are 35 and 4.82 \SI{}\USDMWh{}. Electric demand 2 and generator 1 receive payments to, respectively, cover their bid and generation costs, while generator 2 is charged 3.76 \SI{}\USDMWh{} to cover the payments. Even though generator 1 is charged, its electric surplus increased from zero to \$86.33.

Note that there is no need for uplifts from the heating users and generators, since their bids were higher than the corrected heating price $\gamma^\text{PM}$ of 4.82 \SI{}\USDMWh{}.

A pricing mechanism in which the net generation profit is considered, instead of independent electricity and heating profits, could be done by replacing \eqref{Eq: non-confiscation g} with
\(0 \le \Pi_g + \Theta_g, ~ \forall g\).
The pricing with net generation profit reduces the electricity price, while the heating price is twice the obtained value calculated. In this manner, the heating demand helps cover the economic losses from electricity generation.
Therefore, profit confiscation is presented.
Thus, the proposed cost allocation must be kept separated for each energy vector to avoid cross-subsidies between heat and electricity users.

\subsection{Winter Scenario \label{Sec: Winter}}
Table \ref{Tab: IHPD Results_winter} shows the dispatch results for the winter case. The electricity pricing methodology results are given in Table \ref{Tab: CA Results_winter}.
The electricity and heating prices from the IHPD are 10.95 and 50 \SI{}\USDMWh{}, respectively.
As seen in Fig. \ref{Fig: Results} and Fig. \ref{Fig: EHResults}, none of the generators recovers their marginal electricity costs; incurring in marginal deficits of 35.56 and 19.85 \SI{}\USDMWh{}.

\begin{table}[h]
\caption{Winter IHPD Results\label{Tab: IHPD Results_winter}}\centering
\resizebox{\linewidth}{!}{
\begin{tabular}{@{}ccccc@{}}
\toprule
&\multicolumn{4}{c}{\(SW\) = \$ 5 584.79 
~ \(\lambda^\text{}\) = 10.95 \SI{}\USDMWh{}            
~ \(\gamma^\text{}\) = 50 \SI{}\USDMWh{}}        
 \\ \midrule
User      & Type                                                             & \begin{tabular}[c]{@{}c@{}}Dispatched\\ demand  \(\left[\SI{}\MWh{}\right]\)\end{tabular} & \begin{tabular}[c]{@{}c@{}}Electric\\ surplus  \(\left[\$\right]\)\end{tabular} & \begin{tabular}[c]{@{}c@{}}Heat\\ surplus \(\left[\$\right]\)\end{tabular} 
 \\ \midrule
1 & \(\D\) & 100 & 2 405 & --  \\
2 & \(\D\) & 70 & 1 333.5 & -- \\
3 & \(\Q\) & 164.5 & -- & 0  \\
4 & \(\Q\) & 0 & -- & 0  \\ \midrule 
Generator 
& \begin{tabular}[c]{@{}c@{}}Dispatched\\ electricity \(\left[\SI{}\MWh{}\right]\)\end{tabular} 
& \begin{tabular}[c]{@{}c@{}}Dispatched\\ heat \(\left[\SI{}\MWh{}\right]\)\end{tabular}   
& \begin{tabular}[c]{@{}c@{}}Electric\\ surplus  \(\left[\$\right]\)\end{tabular} 
& \begin{tabular}[c]{@{}c@{}}Heat\\ surplus  \(\left[\$\right]\)\end{tabular} \\
1 & 104.38 & 130.58  & -3 712.51 & 5 379.90 \\
2 & 65.62 & 33.92 &- 1 302.82 & 1 481.58 \\ 
\bottomrule
\end{tabular}
}
\end{table}
%
\begin{table}[h]
\caption{Winter PM--IHPD Results\label{Tab: CA Results_winter}}\centering
\resizebox{\linewidth}{!}{
\begin{tabular}{@{}ccccccc@{}}
\toprule
 \multicolumn{7}{c}{\(SW\) = \$6 934.31 
~ \(\lambda^\text{PM}\) = 45 \SI{}\USDMWh{} 
~ \(\gamma^\text{PM}\) = 50 \SI{}\USDMWh{}} \\ \midrule
User      
& \begin{tabular}[c]{@{}c@{}}\(u^\text{pd}\)  \\ \(\left[\SI{}\USDMWh{}\right]\)\end{tabular}      
& \begin{tabular}[c]{@{}c@{}}\(u^\text{cd}\)  \\ \(\left[\SI{}\USDMWh{}\right]\)\end{tabular}      
& \begin{tabular}[c]{@{}c@{}}\(v^\text{pd}\)  \\ \(\left[\SI{}\USDMWh{}\right]\)\end{tabular}      
& \begin{tabular}[c]{@{}c@{}}\(v^\text{cd}\)  \\ \(\left[\SI{}\USDMWh{}\right]\)\end{tabular} 
& \begin{tabular}[c]{@{}c@{}}Electric\\ surplus \(\left[\$\right]\)\end{tabular}
& \begin{tabular}[c]{@{}c@{}}Heat\\ surplus \(\left[\$\right]\)\end{tabular} \\ \midrule
1 & 0 & 0 & -- & -- & 0 & -- \\
2 & 10 & 0 & -- & -- & 0 & -- \\
3 & -- & -- & 0 & 0 & -- & 0 \\
4 & -- & -- & 0 & 0 & -- & 0 \\ \midrule
Generator 
& \begin{tabular}[c]{@{}c@{}}\(u^\text{pg}\)  \\ \(\left[\SI{}\USDMWh{}\right]\)\end{tabular}      
& \begin{tabular}[c]{@{}c@{}}\(u^\text{cg}\)  \\ \(\left[\SI{}\USDMWh{}\right]\)\end{tabular}      
& \begin{tabular}[c]{@{}c@{}}\(v^\text{pg}\)  \\ \(\left[\SI{}\USDMWh{}\right]\)\end{tabular}      
& \begin{tabular}[c]{@{}c@{}}\(v^\text{cg}\)  \\ \(\left[\SI{}\USDMWh{}\right]\)\end{tabular} 
& \begin{tabular}[c]{@{}c@{}}Electric\\ surplus \(\left[\$\right]\)\end{tabular} 
& \begin{tabular}[c]{@{}c@{}}Heat\\ surplus \(\left[\$\right]\)\end{tabular} \\
1 & 1.52 & 0 & 0 & 0 & 0 & 5 379.90 \\
2 & 0 & 13.08 & 0 & 0 & 72.98 & 1 481.58 \\
\bottomrule
\end{tabular}
}
\end{table}
%

The corrected electricity price is 45 \SI{}\USDMWh{}, Fig. \ref{Fig: EResults}, matching the bid of the electric user 1.
Hence, there is no payment or charge to this user. 
As in the Summer case, uplift payments are made to user 2 and generator 1. 
Generator 2 is charged with an uplift, but it goes from not recovering its costs to having a positive profit.  
Heat uplifts are not necessary, since both generators recover their marginal heat generation costs with the original prices.

\section{Conclusions}
Integrated energy systems are increasingly studied for the enhancement of energy systems' flexibility and cost reduction.
We presented a linear programming methodology for pricing in integrated heat and power markets, guaranteeing cost recovery for the dispatched generators. 
The use of cost recovery per energy resource ensures the avoidance of cross-subsidies between electricity and heating users.
Even though the cost allocation model was designed for a unique node and time step, it can be easily extended to include network and temporal constraints, as well as to integrate other energy infrastructures.

\section*{References}
\bibliographystyle{IEEEtran}
\bibliography{references}

\begin{thebibliography}{1}
\providecommand{\url}[1]{#1}
\csname url@samestyle\endcsname
\providecommand{\newblock}{\relax}
\providecommand{\bibinfo}[2]{#2}
\providecommand{\BIBentrySTDinterwordspacing}{\spaceskip=0pt\relax}
\providecommand{\BIBentryALTinterwordstretchfactor}{4}
\providecommand{\BIBentryALTinterwordspacing}{\spaceskip=\fontdimen2\font plus
\BIBentryALTinterwordstretchfactor\fontdimen3\font minus
  \fontdimen4\font\relax}
\providecommand{\BIBforeignlanguage}[2]{{%
\expandafter\ifx\csname l@#1\endcsname\relax
\typeout{** WARNING: IEEEtran.bst: No hyphenation pattern has been}%
\typeout{** loaded for the language `#1'. Using the pattern for}%
\typeout{** the default language instead.}%
\else
\language=\csname l@#1\endcsname
\fi
#2}}
\providecommand{\BIBdecl}{\relax}
\BIBdecl

\bibitem{Averfalk2017}
H.~Averfalk, P.~Ingvarsson, U.~Persson \emph{et~al.},
  ``\BIBforeignlanguage{en}{Large heat pumps in {{Swedish}} district heating
  systems},'' \emph{\BIBforeignlanguage{en}{Renewable and Sustainable Energy
  Reviews}}, vol.~79, pp. 1275--1284, Nov. 2017.

\bibitem{Guelpa2019}
E.~Guelpa, A.~Bischi, V.~Verda \emph{et~al.}, ``Towards future infrastructures
  for sustainable multi-energy systems: {{A}} review,'' \emph{Energy}, vol.
  184, pp. 2--21, Oct. 2019.

\bibitem{Mitridati2020}
L.~Mitridati, J.~Kazempour, and P.~Pinson, ``\BIBforeignlanguage{en}{Heat and
  electricity market coordination: {{A}} scalable complementarity approach},''
  \emph{\BIBforeignlanguage{en}{European Journal of Operational Research}},
  vol. 283, no.~3, pp. 1107--1123, Jun. 2020.

\bibitem{Cao2019}
Y.~Cao, W.~Wei, L.~Wu \emph{et~al.}, ``Decentralized {{Operation}} of
  {{Interdependent Power Distribution Network}} and {{District Heating
  Network}}: {{A Market}}-{{Driven Approach}},'' \emph{IEEE Trans. Smart Grid},
  vol.~10, no.~5, pp. 5374--5385, Sep. 2019.

\bibitem{Chen2019}
Y.~Chen, W.~Wei, F.~Liu \emph{et~al.}, ``Energy {{Trading}} and {{Market
  Equilibrium}} in {{Integrated Heat}}-{{Power Distribution Systems}},''
  \emph{IEEE Transactions on Smart Grid}, vol.~10, no.~4, pp. 4080--4094, Jul.
  2019.

\bibitem{Schwele2020}
A.~Schwele, A.~Arrigo, C.~Vervaeren \emph{et~al.},
  ``\BIBforeignlanguage{English}{Coordination of {{Electricity}}, {{Heat}}, and
  {{Natural Gas Systems Accounting}} for {{Network Flexibility}}},'' in
  \emph{\BIBforeignlanguage{English}{Proceedings of {{Power Systems Computation
  Conference}}}}, 2020.

\bibitem{Deng2019}
L.~Deng, Z.~Li, H.~Sun \emph{et~al.}, ``Generalized {{Locational Marginal
  Pricing}} in a {{Heat}}-and-{{Electricity}}-{{Integrated Market}},''
  \emph{IEEE Trans. Smart Grid}, vol.~10, no.~6, pp. 6414--6425, Nov. 2019.

\bibitem{Guo1996}
T.~Guo, M.~I. Henwood, and M.~{van Ooijen}, ``An algorithm for combined heat
  and power economic dispatch,'' \emph{IEEE Transactions on Power Systems},
  vol.~11, no.~4, pp. 1778--1784, Nov. 1996.

\bibitem{ONeill2017}
R.~P. O'Neill, A.~Castillo, B.~Eldridge \emph{et~al.}, ``Dual {{Pricing
  Algorithm}} in {{ISO Markets}},'' \emph{IEEE Trans. Power Syst.}, vol.~32,
  no.~4, pp. 3308--3310, Jul. 2017.

\end{thebibliography}

\end{document}